\documentclass{article}
\usepackage{ijcai25}

\usepackage{times}
\usepackage{soul}
\usepackage{url}
\usepackage[hidelinks]{hyperref}
\usepackage[utf8]{inputenc}
\usepackage[small]{caption}
\usepackage{graphicx}
\usepackage{amsmath}
\usepackage{amsthm}
\usepackage{amssymb,bbding}
\usepackage{booktabs}
\usepackage{algorithm}
\usepackage{algorithmic}
\usepackage{multicol,multirow}
\usepackage{adjustbox}
\usepackage{threeparttable}
\usepackage{soul}
\usepackage{color, xcolor} 
\usepackage{subfig}
\usepackage{stfloats}
\usepackage[switch]{lineno}

\title{Emerging Advances in Learned Video Compression: Models, Systems and Beyond}

\author{
Chuanmin Jia$^1$
\and
Feng Ye$^1$\and
Siwei Ma$^{1,2}$\and
Wen Gao$^{1,2}$\and
Huifang Sun$^{2,3}$\And
Leonardo Chiariglione$^{2,3}$\\
\affiliations
$^1$Peking University,
$^2$Peng Cheng Laboratory, $^3$MPAI
}

\begin{document}

\maketitle

\begin{abstract}
    %Video compression is a fundamental topic in the visual information processing chain, linking visual sensing/capturing and high-level visual analytics. Numerous lossy video compression algorithms and standards are found in existing literature. In recent years, the broad success of deep learning has enriched the horizon of video compression into novel paradigms by leveraging end-to-end optimized neural models. In this survey, we first provide a comprehensive and systematic overview of recent literature on end-to-end optimized learned video coding, covering the spectrum of pioneering efforts in compression model designation and rate-distortion efficiency analysis, including neural network based predictive coding approaches, encoder-decoder architectures for motion field and residual compression, multi-scale hierarchical prediction structure, recurrent networks, visual attention models, flow-based models, and generative models in adversarial settings. Then, we further investigate the system design and hardware implementation challenges of the learned video codecs inclusively and discuss the coding principles behind these models. Finally, we make efforts to answer the question that why learned codecs would be the promising direction of future research.
    Video compression is a fundamental topic in the visual intelligence, bridging visual signal sensing/capturing and high-level visual analytics. The broad success of artificial intelligence (AI) technology has enriched the horizon of video compression into novel paradigms by leveraging end-to-end optimized neural models. In this survey, we first provide a comprehensive and systematic overview of recent literature on end-to-end optimized learned video coding, covering the spectrum of pioneering efforts in both uni-directional and bi-directional prediction based compression model designation. We further delve into the optimization techniques employed in learned video compression (LVC), emphasizing their technical innovations, advantages. Some standardization progress is also reported. Furthermore, we investigate the system design and hardware implementation challenges of the LVC inclusively. Finally, we present the extensive simulation results to demonstrate the superior compression performance of LVC models, addressing the question that why learned codecs and AI-based video technology would have with broad impact on future visual intelligence research.
\end{abstract}

\section{Introduction}

Video data has emerged as the largest category of big data in visual information processing, with its volume continuing to increase exponentially. The majority of global network traffic has been dominated by video applications with the growing number of Internet users. Consequently, advancements in image and video compression technologies have become critical, as they directly impact over half of the world’s population. Driven by emerging functional requirements and application scenarios such as interactive video conferencing and remote screen sharing, there is an increasing need for advanced coding techniques to ensure efficient video transmission, storage, and enhanced quality-of-experiences. Lossy video compression plays a fundamental role in handling these demands. %It can be formulated as an optimization problem aimed at reconstructing video signals with maximal quality and minimum coding length (entropy) under the condition of compressed data representation, which originates from the lossy source coding principle of information theory.

Numerous video compression approaches have been investigated, enabling plenty of video services and standardized contributions~\cite{bross2021overview,ma2022evolution}. More recently, deep neural networks have significantly advanced various fields, and their application to video coding has garnered increasing attention for their remarkable non-linear modeling ability to enrich the horizon of image and video coding~\cite{ma2019image,liu2020deep}. 
%In this context, end-to-end learned video compression (LVC) has been an emerging research topic due to its overall optimization capability and promising coding efficiency, also enabling video representation learning for ubiquitous multimedia. 
%A variety of LVC algorithms have been developed in recent literature, ranging from framework-level innovations to algorithm-level advancements. In addition, the open-source models significantly accelerate the iterative process of LVC, rapidly increasing the coding performance. 
Evidence could be observed that they achieve promising rate-distortion (R-D) performances on commonly adopted test benchmarks and realize comparable performances with the state-of-the-art standardized coding methods. 
%This has caused a paradigm shift in the video coding field. 
Given its substantial R-D performance, flexible technological development, plug-and-play hardware deployment, and adaptability to future requirements in intelligent media applications, LVC is poised to be the next major advancement in video compression.

Apart from the academic research on LVC, efforts are also made from the standardization perspective. The Moving Picture Experts Group (MPEG) and Joint Video Expert Team (JVET) have been continuously studying the requirements and technological tools for neural coding solutions since 2019. Modular neural network-based coding tools provide sufficient coding gain for almost every single module of the hybrid framework, resulting in LVC by replacing existing tools with neural approaches. The Moving Picture, Audio and Data Coding by Artificial Intelligence (MPAI) organization has established a novel project named end-to-end video coding (EEV) in early 2022, targeting at longer-term needs for fully neural-network-based video compression. The Audio and Video coding Standard Group (AVS) has also initiated End-to-end video coding Exploration Model (EEM) to explore low latency and computational complexity compression models.

This paper aims to provide a \textit{comprehensive, systematic, and up-to-date survey} on LVC, ranging from model formulation to merging methods and applications. It is believed that the survey can serve as an innovative starting point for understanding the development situation of LVC field, identifying the limitations of existing methods, and exploring potential future directions. To this end, we begin with providing an overview of the existing lossy LVC model architectures and their evolution history, which serve as the backbones of this survey in Section 2. Section 3 discusses key optimization techniques relevant to LVC, while Section 4 elaborates on the system design and hardware implementation. In Section 5, we conduct experimental evaluations to demonstrate the compression performance of state-of-the-art LVC models. Finally, we present our conclusions in section 6. We deeply hope this survey provides novel insights for promoting the advanced LVC model both for the machine learning and signal processing community.

\section{Overview of LVC}
\begin{figure*}[tb]
  \centering
  \footnotesize
  \subfloat[Traditional video codec]{
    \includegraphics[height=4cm]{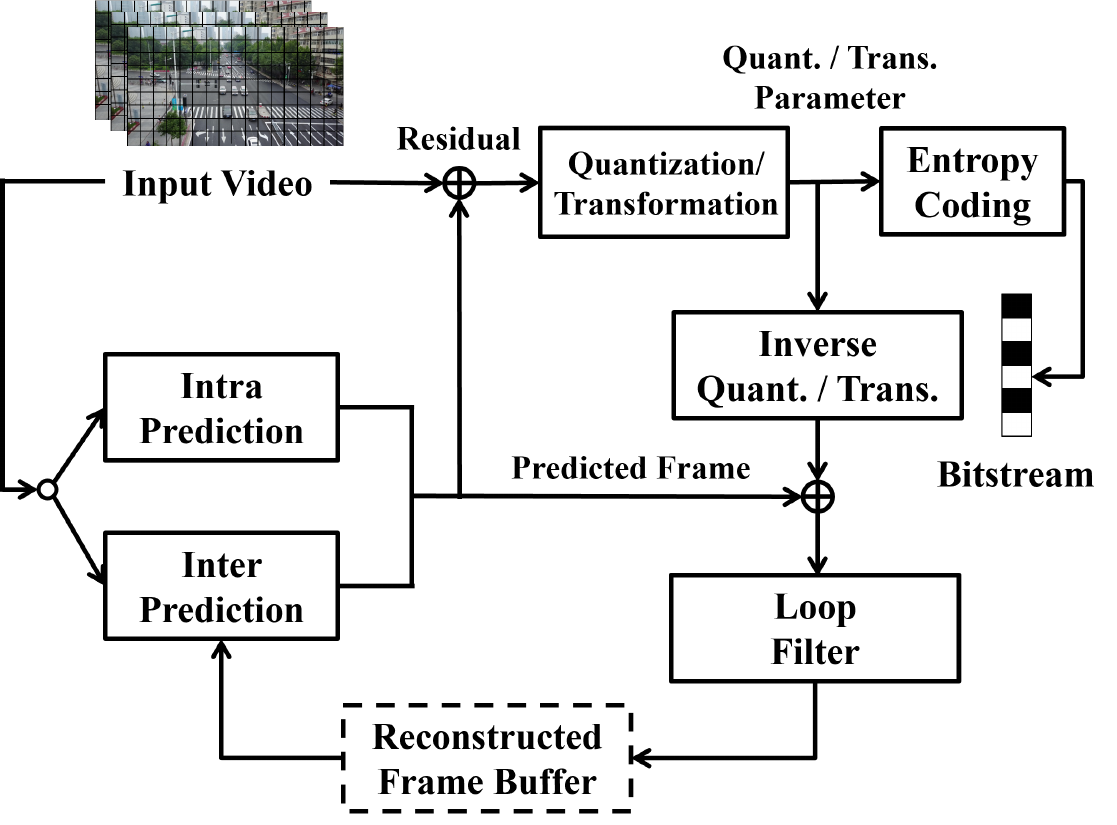}
    \label{fig:framework1}}\hfil
  \subfloat[Uni-directional prediction based LVC]{
    \includegraphics[height=4.4cm]{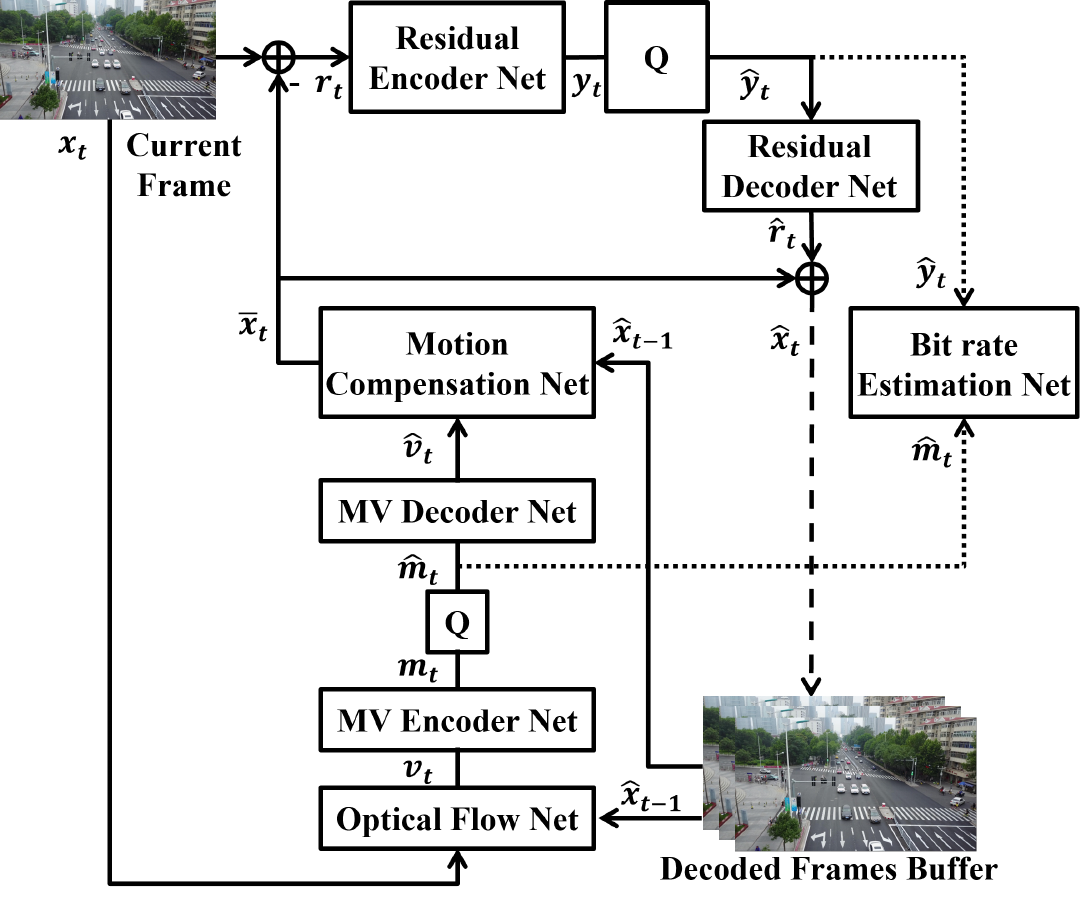}
    \label{fig:framework2}}\hfil
  \subfloat[Bi-directional prediction based LVC]{
    \includegraphics[height=4.45cm]{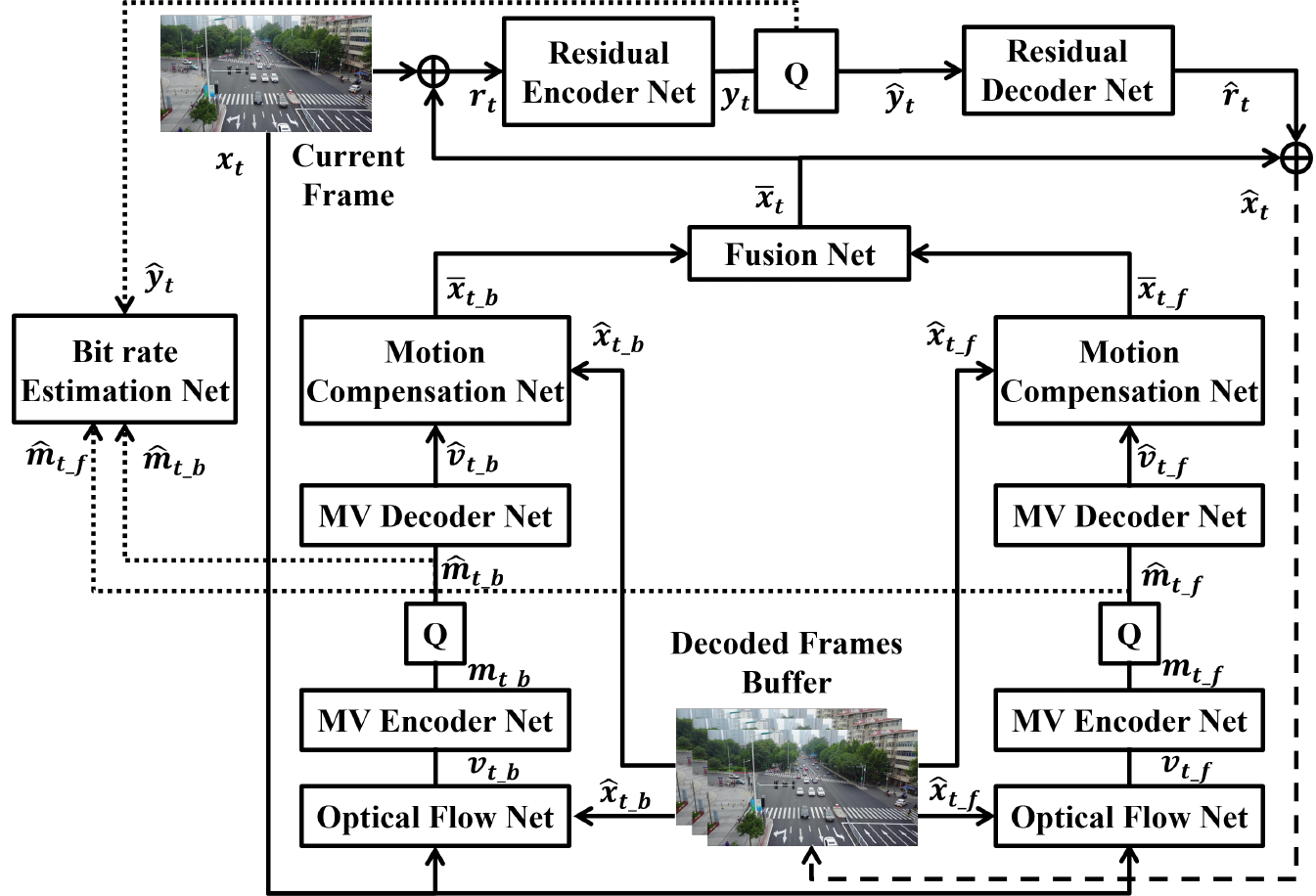}
    \label{fig:framework3}}
%\captionsetup{font=small}
  \caption{The direct architecture comparisons among different video compression methods. (a) Conventional hybrid video codec, (b) Uni-directional prediction based LVC. (c) Bi-directional prediction based LVC. Conventional codec suffers from local-modular optimization with marginal efficiency improvement. The latter two structures support end-to-end (E2E) optimization.}
  \label{fig:framework}
\end{figure*}

As depicted in Fig.\ref{fig:framework1}, existing traditional video coding standards employ a block-based transform-prediction hybrid coding framework. Frames are partitioned into variable-sized blocks, with each encoded using optimal intra-prediction modes based on reconstructed neighboring blocks (left and above) for spatial context. Inter-frame prediction further enhances efficiency through motion estimation and compensation, identifying similar blocks as hypotheses. Such an approach is inherently block-dependent and requires sequential processing, which limits its flexibility and efficiency.

In contrast, LVC leveraging neural networks exhibits superior nonlinear modeling capabilities and enables end-to-end joint optimization of all modules, thereby achieving significantly higher overall performance. This advantage allows LVC frameworks to incorporate more sophisticated modules and intricate designs as shown in Fig.\ref{fig:framework2} and \ref{fig:framework3}. Pioneering research in LVC, such as the works presented in ~\cite{chen2017deepcoder,lu2019dvc}, laid the foundation for this paradigm shift. To provide a clear and structured overview of the advancements in this field, we systematically introduce the frameworks and technical evolution of typical end-to-end optimized LVC models. The discussion is organized into two categories: uni-directional and bi-directional prediction-based methods, highlighting their respective developments and contributions to the field.  Summaries of uni-directional and bi-directional predicted LVC models are shown in Table \ref{tab:uni-directional} and \ref{tab:bi-directional}.

% 要加一下为什么要区分P帧和B帧
% 需要铺垫 通过事实去引出 为什么要这么写

\begin{table*}[]
\caption{A comprehensive model summary of uni-directional predicted (P-frame based) LVC models. The modulized feature and network design are comparatively illustrated to better understand the design philosophy behind those models.}\footnotesize
\begin{adjustbox}{center}
\begin{threeparttable}
\begin{tabular}{c|ccccccccc}
\hline\hline
\textbf{Methodology} & \textbf{\begin{tabular}[c]{@{}c@{}}Motion Estimation\\Network\end{tabular}} & \textbf{\begin{tabular}[c]{@{}c@{}}Motion Compensation\\Network\end{tabular}} & \textbf{\begin{tabular}[c]{@{}c@{}}Coding\\Manner\end{tabular}} & \textbf{\begin{tabular}[c]{@{}c@{}}Feature\\Extractor\end{tabular}} & \textbf{\begin{tabular}[c]{@{}c@{}}MRF\end{tabular}} & \textbf{\begin{tabular}[c]{@{}c@{}}VBSM\end{tabular}} & \textbf{\begin{tabular}[c]{@{}c@{}}MMC\end{tabular}}   \\ \hline\hline
%DVC\shortcite{lu2019dvc}              & \Checkmark                                            & Pixel                                              & Residual                                          & \XSolidBrush                                          & \XSolidBrush                                                               & \XSolidBrush                                                           & \XSolidBrush                                                           \\
%M-LVC\shortcite{lin2020m}              & \Checkmark                                            & Pixel                                              & Residual                                          & \XSolidBrush                                         & \Checkmark                                                            & \XSolidBrush                                                        & \XSolidBrush                                                            \\
RaFC~\shortcite{hu2020improving}              & \Checkmark                                            & Pixel                                              & Residual                                          & \XSolidBrush                                          & \XSolidBrush                                                            & \XSolidBrush                                                        & \XSolidBrush                                                           \\
NVC~\shortcite{liu2020neural}              & \Checkmark                                            & Feature                                              & Residual                                          & \XSolidBrush                                          & \XSolidBrush                                                            & \XSolidBrush                                                        & \Checkmark                                                            \\
DVCPro~\shortcite{9072487}              & \Checkmark                                            & Pixel                                              & Residual                                          & \XSolidBrush                                          & \XSolidBrush                                                               & \XSolidBrush                                                           & \XSolidBrush                                                            \\
%RLVC\shortcite{Yang_2021}              & \Checkmark                                            & Pixel                                              & Residual                                          & \XSolidBrush                                          & \XSolidBrush                                                            & \XSolidBrush                                                        & \XSolidBrush                                                            \\
DCVC~\shortcite{li2021deep}              & \Checkmark                                            & Feature                                              & Conditional                                          & \Checkmark                                          & \XSolidBrush                                                            & \XSolidBrush                                                        & \XSolidBrush                                                           \\
%FVC\shortcite{hu2021fvc}              & \Checkmark                                            & Feature                                              & Residual                                          & \Checkmark                                          & \Checkmark                                                            & \XSolidBrush                                                        & \XSolidBrush                                                            \\
ELF-VC~\shortcite{rippel2021elf}              & \Checkmark                                            & Pixel                                              & Residual                                          & \XSolidBrush                                          & \XSolidBrush                                                            & \Checkmark                                                        & \XSolidBrush                                                            \\
VLVC(lowdelay)~\shortcite{Feng2021VersatileLV}              & \Checkmark                                            & Pixel                                              & Residual                                          & \XSolidBrush                                          & \Checkmark                                                            & \XSolidBrush                                                        & \XSolidBrush                                                            \\
%DCVC-TCM\shortcite{sheng2021temporal}              & \Checkmark                                            & Feature                                              & Conditional                                          & \Checkmark                                          & \XSolidBrush                                                            & \XSolidBrush                                                        & \Checkmark                                                            \\
DCVC-HEM~\shortcite{Li_2022}              & \Checkmark                                            & Feature                                              & Conditional                                          & \Checkmark                                          & \XSolidBrush                                                            & \Checkmark                                                        & \Checkmark                                                           \\
C2F-DVC~\shortcite{9880063}              & \Checkmark                                            & Feature                                              & Residual                                          & \Checkmark                                          & \XSolidBrush                                                            & \XSolidBrush                                                        & \Checkmark                                                            \\
CANF-VC~\shortcite{Ho2022CANFVCCA}              & \Checkmark                                            & Pixel                                              & Conditional                                          & \XSolidBrush                                          & \Checkmark                                                            & \XSolidBrush                                                        & \XSolidBrush                                                           \\
Alpha-VC~\shortcite{Shi2022AlphaVCHA}              & \XSolidBrush                                            & -                                              & Residual                                          & \Checkmark                                          & \XSolidBrush                                                            & \XSolidBrush                                                        & \XSolidBrush                                                           \\
MPAI EEV-0.4~\shortcite{jia2023mpai}              & \Checkmark                                            & Feature                                              & Residual                                          & \Checkmark                                          & \XSolidBrush                                                            & \XSolidBrush                                                        & \XSolidBrush                                                            \\ 
%DCVC-DC\shortcite{li2023neural}              & \Checkmark                                            & Feature                                              & Conditional                                          & \Checkmark                                          & \XSolidBrush                                                            & \Checkmark                                                        & \Checkmark                                                           \\
DCVC-FM~\shortcite{li2024neural}              & \Checkmark                                            & Feature                                              & Conditional                                          & \Checkmark                                          & \XSolidBrush                                                            & \Checkmark                                                        & \Checkmark                                                           \\
DCVC-SDD~\shortcite{sheng2024spatial}              & \Checkmark                                            & Feature                                              & Conditional                                          & \Checkmark                                          & \XSolidBrush                                                            & \Checkmark                                                        & \Checkmark                                                            \\
\hline\hline
\end{tabular}
\begin{tablenotes}
    \footnotesize
    \item It should be noted that the ``VBSM'' refers to Variable Bitrate in a Single Model, ``MMC'' refers to Multi-scale Motion Compensation. ``MRF'' refers to Multiple Reference Frames/Features, respectively.
  \end{tablenotes}
\end{threeparttable}
\end{adjustbox}
\label{tab:uni-directional}
\end{table*}

\subsection{Uni-directional Prediction based LVC} 

% \begin{figure}[tb]
%   \centering
%     \includegraphics[height=2.3cm]{uni-roadmap.png}
%   %\fbox{\rule[-.5cm]{0cm}{4cm} \rule[-.5cm]{4cm}{0cm}}
% %\captionsetup{font=small}
%   \caption{The technical milestones of Uni-directional Prediction based LVC.}
%   \label{fig:uni-roadmap}
% \end{figure}

%{\bf Pixel Space Prediction.}  ,lu2020content
Since 2017, primary explorations on LVC have initially focused on pixel-domain operations to reduce redundancy. Early works such as ~\cite{chen2017deepcoder,lu2019dvc} utilize pixel-level optical flow for motion estimation and motion compensation to compress residuals. These methods employ complex structural designs, such as pre-trained optical flow networks and complicated training schemes. Therefore, pixel-domain operations have inherent limitations. First, current optical flow networks struggle to generate accurate pixel-level optical flow information, especially for high-resolution videos with complex motion. Second, pixel-domain operations may introduce significant artifacts during motion compensation. Furthermore, producing reliable pixel-level residual information is challenging, and compressing it is inherently difficult. Despite these limitations, these early models lay the foundation for end-to-end optimized video codecs, and their shortcomings in model design and structure organization spur rapid advancements in subsequent research.

%{\bf Feature Space Prediction.}
To address the limitations of pixel-level residuals, researchers shifted their focus to feature-domain operations. In 2020, a significant advancement in this direction involves calculating residuals in high-dimensional feature space and transforming them into discrete latent representations, rather than directly computing pixel-level residuals, as demonstrated in~\cite{Feng2020LearnedVC}. This approach marked a notable departure from traditional pixel-domain methods. Building on this progress, a feature space video coding network was developed to perform all operations including motion estimation, motion compensation, and residual compression in the feature domain, as shown in ~\cite{hu2021fvc}. By fusing multiple reference features stored in the decoding buffer to reconstruct frames, this method achieved superior compression performance. Further advancements include learning temporal contexts from propagated features rather than previously reconstructed frames, as proposed in ~\cite{sheng2021temporal}, which enhanced system performance by capturing more meaningful temporal information during encoding.

%{\bf Conditional Coding.} 
Residual coding has been a cornerstone of LVC methods, particularly before 2020, due to its ability to effectively leverage the strong temporal correlations between frames. By encoding the difference between the current frame and its prediction from reference frames, residual coding has proven to be a practical approach. However, Shannon entropy analysis reveals that the entropy of residual coding is always greater than or equal to that of conditional coding, suggesting that residual coding is not always the most efficient choice. This limitation prompted the exploration of alternative methods, leading to the development of conditional coding.

Conditional coding emerged as a more extensible and efficient alternative to residual coding, defining conditions as learnable feature-domain contexts that can incorporate rich information to aid in encoding, decoding, and entropy modeling. Initially applied to specific modules such as entropy and foreground content coding, as demonstrated in ~\cite{ladune2020optical}, conditional coding was later expanded into a comprehensive framework. This framework utilized motion estimation and motion compensation in the feature domain to generate contextual features, significantly enhancing the capacity of conditional information, as shown in ~\cite{li2021deep}. The potential of conditional coding was further realized in ~\cite{Ho2022CANFVCCA}, where it was extended to motion and integrated into a fully conditional coding-based video compression framework, achieving state-of-the-art performance and highlighting its superior advantages over traditional residual coding methods.

%{\bf Multiple Reference Pictures/Features.} 
While the previous discussion focused on single-reference-frame scenarios, the use of multiple reference frames or features has been shown to enhance prediction performance by providing richer motion information and improving motion estimation and compensation accuracy. Compared to frameworks that rely on a single reference frame, multiple reference frames can significantly reduce the residual or conditional information that needs to be encoded, thereby lowering the bit rate and improving coding efficiency. For instance, the use of multiple previous frames as references was proposed to generate more accurate predictions and minimize residual information, accompanied by a motion vector (MV) prediction module to extrapolate and predict MVs, further reducing coding costs ~\cite{lin2020m}. Additionally, multiple reference frames help the encoder adapt to diverse scenes and content, such as sequences with stationary backgrounds, complex motion, or occlusion, where they can deliver substantial performance gains. A multiple feature fusion module based on a non-local attention mechanism was introduced to leverage features from multiple previous frames, aiding in reconstructing the current frame ~\cite{hu2021fvc}. This concept was further advanced by designing a flow extrapolation network to generate effective conditions for conditional motion coding, demonstrating the significant advantages of using multiple reference frames and features ~\cite{Ho2022CANFVCCA}.

%{\bf Frame Generation/Forecasting.} 
%While leveraging multiple reference frames has proven effective in enhancing prediction performance, another promising direction lies in exploiting the strong temporal correlations between adjacent frames through deep generative modeling. Researchers have explored this approach to efficiently code latent variables associated with each frame, achieving significant improvements in compression efficiency. For example, a time-conditional prior distribution parameterized by a deep generative model was proposed to encode latent variables, resulting in smaller entropy and code rates ~\cite{Lombardo2018DeepGV}. Further advancements were made by interpreting video compression as an instantiation of autoregressive flow transformation, incorporating insights from generative modeling to enhance performance ~\cite{yang2021generative}. Building on these developments, an end-to-end learning video compression codec based on GAN-prior generative image compression was introduced, utilizing a frame generation network-based autoencoder architecture to achieve superior synthesis and coding performance ~\cite{9578733}. These innovations underscore the potential of deep generative modeling in pushing the boundaries of video compression efficiency.

In the rapidly evolving field of P-frame-based LVC, state-of-the-art models have demonstrated superior rate-distortion (R-D) performance compared to conventional video codecs. This progress highlights the potential of LVC for practical applications. Future research is expected to focus on improving robustness and developing content-dependent coding methods, as network structures continue to converge and mature.

\subsection{Bi-directional Prediction based LVC}
While bi-directional prediction-based methods in LVC have received less attention compared to uni-directional prediction approaches and face challenges in achieving comparable performance, they still represent a significant and evolving area of research with their own unique development trajectory and contributions. This section explores the advancements in bi-directional prediction-based LVC, focusing on two key strategies such as frame interpolation and hierarchical coding structures.

\begin{table}[]
\centering
\caption{A summary of Bi-directional predicted (B-frame based) LVC models. The modulized feature and network design are comparatively illustrated to better understand the design philosophy behind those models.}\tiny
\begin{adjustbox}{center}
\begin{tabular}{c|ccccc}
\hline\hline
\textbf{Methodology} & \textbf{\begin{tabular}[c]{@{}c@{}}Motion\\coding\end{tabular}} & \textbf{\begin{tabular}[c]{@{}c@{}}Frame\\Interpolator\end{tabular}} & \textbf{\begin{tabular}[c]{@{}c@{}}Coding\\Manner\end{tabular}} & \textbf{\begin{tabular}[c]{@{}c@{}}Hierarchical\\Coding\\Structure\end{tabular}} & \textbf{\begin{tabular}[c]{@{}c@{}}Feature\\Extractor\end{tabular}}  \\ \hline\hline
Compress AE~\shortcite{wu2018video}              & \Checkmark                                              & \Checkmark                                                        & Residual                                          & Hand-crafted                                          & \XSolidBrush                                                            \\
NeuralInter~\shortcite{djelouah2019neural}              & \Checkmark                                              & \Checkmark                                          & Residual                                                            & Self-adaption                                                        & \XSolidBrush                                                     \\
HLVC~\shortcite{yang2020learning}              & \Checkmark                                              & \XSolidBrush                                                        & Residual                                          & Hand-crafted                                          & \XSolidBrush                                                            \\
LHBDC~\shortcite{yilmaz2021end}              & \Checkmark                                              & \XSolidBrush                                          & Residual                                                            & Self-adaption                                                        & \XSolidBrush                                                     \\
B-EPIC~\shortcite{pourreza2021extending}              & \Checkmark                                              & \Checkmark                                          & Residual                                                            & Self-adaption                                                        & \XSolidBrush                                                     \\ 
Conditional~\shortcite{Ladune2021ConditionalCF}              & \Checkmark                                              & \XSolidBrush                                          & Conditional                                                            & Self-adaption                                                        & \XSolidBrush                                                     \\
VLVC~\shortcite{Feng2021VersatileLV}              & \Checkmark                                              & \XSolidBrush                                                        & Residual                                          & Self-adaption                                          & \XSolidBrush                                                            \\
B-CANF~\shortcite{Chen2022BCANFAB}              & \Checkmark                                              & \XSolidBrush                                          & Conditional                                                            & Self-adaption                                                           & \XSolidBrush                                                     \\
TLZMC~\shortcite{Alexandre2023HierarchicalBV}              & \XSolidBrush                                              & \Checkmark                                          & Conditional                                                            & Self-adaption                                                        & \XSolidBrush                                                     \\ 
Bi/Hi-DCVC~\shortcite{Kim2023NeuralVC}              & \Checkmark                                              & \XSolidBrush                                                        & Conditional                                          & Self-adaption                                          & \Checkmark                                                            \\
DVC-SHBMM~\shortcite{deepye2024}              & \Checkmark                                              & \XSolidBrush                                                        & Residual                                          & Self-adaption                                          & \Checkmark                                                            \\
DCVC-B~\shortcite{sheng2024bi}              & \Checkmark                                              & \XSolidBrush                                                        & Conditional                                          & Self-adaption                                          & \Checkmark                                                            \\
\hline\hline
\end{tabular}
\end{adjustbox}
\label{tab:bi-directional}
\end{table}

%{\bf Frame Interpolation.} 
Bi-directional prediction leverages both preceding and subsequent reference frames, making frame interpolation an intuitive and effective method for enhancing compression. Early work by ~\cite{wu2018video} treated video compression as a repetitive image interpolation task, employing deep image generation and interpolation for hierarchical coding of B-frames. Building on this, an interpolation model was proposed by ~\cite{djelouah2019neural} that combined motion compression and image synthesis while reducing computational overhead during decoding. Further advancements were made by ~\cite{pourreza2021extending}, who interpolated two reference frames to generate a single reference frame, which was then used alongside an existing P-frame codec to encode the input B-frame. A more recent innovation by ~\cite{Alexandre2023HierarchicalBV} introduced a low-resolution base layer containing interpolated frames, replacing the motion encoder to improve compression performance while reducing complexity.

%{\bf Hierarchical Coding Structure.} 
In addition to frame interpolation, hierarchical coding structures have also been explored in bidirectional LVC, offering the advantage of adapting to more flexible coding configurations while enhancing compression efficiency. These structures, widely used in traditional video coding standards, allow high-quality frames to serve as superior references, improving the compression of other frames. However, research on hierarchical B-frame schemes in LVC remains limited. In ~\cite{yang2020learning}, a Hierarchical Learned Video Compression (HLVC) method with three quality layers was proposed, applying different coding methods and compression qualities to each layer. While effective, this approach suffered from a fixed layered structure and the use of separate compression models for each layer, leading to increased memory and computational costs. Addressing these limitations, ~\cite{Kim2023NeuralVC} introduced a bidirectional LVC model with a hierarchical structure similar to conventional standards, incorporating temporal layer-adaptive optimization. This approach enabled hierarchical quality regulation within a single model, offering a more efficient and flexible solution.

In 2024, significant advancements were made in bidirectional learned video compression (LVC) models. A novel bidirectional LVC model featuring two parameter-shared motion codecs was introduced in ~\cite{deepye2024}, which effectively leverages scaled motion information from temporal contexts as prior. This model incorporates trustworthy motion modeling and an efficient rate allocation strategy to optimize overall compression performance. DCVC-B was also developed in \cite{sheng2024bi}, which utilized a bi-directional motion difference context propagation method and a contextual compression model. These developments have enabled LVC models to surpass the performance of the H.266/VVC reference software on specific test datasets under same configurations in both RGB and YUV420 color spaces.

In summary, while bidirectional prediction-based LVC methods are still in their developmental stages compared to unidirectional approaches, they have demonstrated promising progress through innovations in frame interpolation and hierarchical coding structures. These advancements highlight the potential of bidirectional prediction to overcome current limitations and contribute to the broader evolution of learned video compression.

% Bi-prediction
% quality enhancement

\section{Optimization Techniques for LVC}
The data-trained LVC models have different plugin-in modules and R-D behaviors compared to conventional codecs. Therefore, the optimization of LVC frameworks remains a critical area of research, as it directly impacts compression performance, adaptability, and practical applicability. This chapter delves into six key optimization methods to highlight their technical innovations, advantages, and contributions to advancing the state-of-the-art in LVC. By systematically examining these optimization strategies, this chapter provides a comprehensive understanding of how LVC frameworks can be fine-tuned to meet diverse compression requirements and application scenarios.

\subsection{Encoder-side Online Learning}
LVC models tend to suffer from severe error propagation due to the accumulation of reconstructed error in low-delay inter-predictive coding. To reduce the error propagation and online adapt the contents, an error propagation aware (EPA) training strategy was proposed in~\cite{lu2020content}. Considered more temporal information to alleviate error accumulation, this method supported inference stage adaptation by allowing the online update of the video encoder without introducing any extra parameters. Specifically, to optimize the encoder for each frame and mitigate the domain gap between training and testing data, the online encoder updating scheme is in the inference stage. The encoder part is updated according to the input image while keeping the decoder unchanged. In contrast to other low-level vision tasks, the pristine video frames are available at the encoder side. As a result, it is practical to update the encoder by using the original frame as long as the decoder remains unchanged.

\subsection{Decoder-side Content Adaptation} Similar to conventional video codecs, the decoder side video content adaptation for quality enhancement significantly improves the reconstructed quality of decoded videos. The author borrowed the knowledge from the quality post-processing models~\cite{yang2020learning} to realize compression artifacts removal and noise reduction. The decoder-side motion derivation was explored in~\cite{chen2022learning} using the LTSM-UNet structure to efficiently capture both spatial and temporal information. In this manner, it is obvious that the computational overhead of motion estimation and motion compensation is avoided. However, using a displacement calculation unit to store and update the motion information results in poor performances in high-bitrate coding.

%\subsection{R-D Characteristics of LVC}
%The data-trained LVC models have different R-D behaviors compared to conventional codecs. Therefore, it is necessary to consider the problem of R-D characteristics and modeling. Several studies describing the R-D analysis of LVC methods are proposed to support multi-rate compression by inserting plugin-in modules to express the relationship between the target bitrate and the latent variables. With variable-rate compression ability, the underlying rate control models can be formulated for R-D characteristic description.

\subsection{Variable Bitrate Encoding}
Variable bitrate encoding in LVC has been achieved through two primary approaches: network modulation and mask modeling for latent features. Network modulation techniques, such as those proposed in ~\cite{lin2021modulated}, use a rate-distortion (R-D) tradeoff parameter to modulate internal feature maps of motion and residual auto-encoders, enabling variable rate adaptation within a single model. Multi-rate training strategies, including multi-rate-distortion loss functions and step-by-step optimization, have further enhanced this capability ~\cite{rippel2021elf}. Additionally, discrete and continuous rate adaptation has been realized through mechanisms like \textit{Gain Units} and exponential interpolation, though interpolation-based methods often face limitations in bitrate range ~\cite{cui2021asymmetric}. On the other hand, mask modeling for latent features, such as the content-adaptive bit allocation method in ~\cite{lin2022content}, dynamically adjusts the quantization strategy through latent scaling. Binarized mask modeling, as explored in ~\cite{fathima2023neural}, offers a flexible and training-stable alternative for adjusting bitrates by simulating different masking ratios of latent features. These approaches collectively enable single-model multi-rate adaptation, either by implicitly smoothing entropy through auxiliary sub-modules or explicitly modeling binarized masks for latent elements.

\subsection{Rate Control Models and Methods}
Rate control in LVC has been addressed through optimization-based and empirical-based methods. Optimization-based approaches, such as Semi-Amortized Variational Inference (SAVI), model rate control as a GoP-level optimization problem, harmonizing parameter updates with the learning process of LVC models to achieve optimal rate control ~\cite{xu2023bit}. Reinforcement learning techniques, including imitation learning within a Partially Observable Markov Decision Process (POMDP) framework, have also been applied to learn neural rate control policies ~\cite{mao2020neural}. Empirical-based methods, such as the deep learning-based rate implementation network in ~\cite{zhang2024neural}, allocate optimal bitrates to frames based on spatial and temporal characteristics. Additionally, sparse-to-dense R-D point conversion schemes, as proposed in ~\cite{chen2023sparse}, enhance rate control flexibility without increasing model complexity. While empirical methods are easier to implement, they often suffer from performance limitations, whereas optimization-based methods, despite their high computational cost, offer more precise control. The choice of rate control method ultimately depends on specific application requirements and scenarios.

\subsection{Perceptual Quality Optimization}
Human perceptual quality is at the core of lossy LVC, with numerous approaches developed for perceptual quality assessment and improvement. The authors of~\cite{pergament2022pim} developed an interactive web-tool that allows annotators to construct spatio-temporal perceptual importance maps from a video.
Regarding the extremely low bitrate coding situation, the learning objective from the adversarial perspective was able to improve the perceptual quality of decoded videos~\cite{veerabadran2020adversarial}. In this context, recurrent conditional generative adversarial learning was introduced in~\cite{yang2021perceptual} to generate photo-realistic and temporally coherent compressed frames. This training strategy ensured a balance between bitrate, distortion and perceptual quality with reasonable Generalizability.

Generally speaking, the perceptual quality-oriented LVC models usually propose hybrid objective functions (signal level plus perceptual level loss) associated with enhanced learning strategies such as discriminative-generative learning, adversarial learning, region-of-interest masking, etc.

%\subsection{Training Strategy}

\subsection{Network Model Quantization}
Model quantization of neural networks is a critical step towards deployment for LVC, ensuring interoperability among different devices especially guaranteeing the entropy decoding of the bitstream. Mainstream model post-training quantization techniques for learned compression models encapsulate weights scalar quantization and calibration, to enable reliable cross-platform encoding and decoding procedures using variational models. To overcome the catastrophic mismatch of the Gaussian prior between the sending and the receiving side, \cite{balle2018integer} proposed to train the networks entirely using floating point computations, then rounded to integers after every computational operation, while the back-propagation was done with full floating point precision. In addition, they pre-computed all possible values and expressed them as a lookup table to ensure the deterministic computation of prior distribution. The range-constrained activation and the fixed-point processing for range-adaptive quantization to convolution kernels were introduced in~\cite{hong2020efficient}, keeping the fixed-point arithmetic with significant complexity reduction and robust performance at the inference stage. 

To realize LVC interoperability, network-weights quantization is mandatory for LVC methods. However, current quantization schemes usually enjoy mix-precision bit-depth and input-dependent, which shows little friendly to the practical utility of LVC. Fixed-bit quantization of weights and activations might be harmful to the R-D efficiency, especially in high-bitrate coding scenarios. Therefore, weight calibration and quantization-aware training should be further applied after coarse quantization process.

The optimization methods discussed in this chapter underscore the versatility and adaptability of LVC frameworks. These advancements not only improve compression efficiency but also enable LVC to cater to a wide range of practical applications, from low-latency streaming to high-quality video storage. As research in LVC continues to evolve, the integration and refinement of these optimization techniques will play a pivotal role in bridging the gap between theoretical innovation and real-world deployment. The insights provided in this chapter lay the groundwork for future explorations, paving the way for more efficient, adaptive, and scalable video compression solutions.

\section{LVC System Design and Implementation}
Towards real-world LVC utility, the system-level design, interface collaboration and hardware implementation (e.g., resource-constrained, edge-computing devices and mobile devices) are crucial parts of LVC research and significantly impact the practical usage and deployment. 
These aspects are particularly vital when considering the diverse hardware environments in which LVC operates, ranging from resource-constrained edge-computing devices to mobile devices, each presenting unique challenges and opportunities. 
%The design and optimization of LVC systems must account for computational efficiency, energy consumption, and adaptability to varying hardware capabilities, ensuring seamless integration into real-world scenarios. 
In this section, we delve into system-level designs and hardware implementation strategies that underpin LVC's practical adoption.

\begin{table*}[tb]
%\captionsetup{font=small}
\caption[]{BD-Rate(\%) performance comparison using PSNR (RGB colorspace). The intra-period is set to be 32. Baseline is HM-16.20.}
\centering
%\footnotesize
\label{tab:1}
\begin{tabular}{c cccccc cc}
\toprule
  & \footnotesize{JCT-VC Class B} & \footnotesize{JCT-VC Class C} & \footnotesize{JCT-VC Class D} & \footnotesize{JCT-VC Class E} & \footnotesize{UVG} & \footnotesize{MCL-JCV} & \footnotesize{\bf Average} \\
\midrule
%HM-16.20       & 0                             & 0                             & 0                             & 0                             & 0                             & 0                             & 0                             \\
%B-CANF & -4.46                         & 9.35                         & -13.33                          & 0.03                         & -2.38                         & 13.48                        & 5.49                        \\
%DCVC-HEM & -8.40                         & 6.40                         & -5.97                          & -3.88                         & -3.24                         & -2.50                        & -7.67                        \\
DCVC-DC~\shortcite{li2023neural}  & -24.54 & -6.12                        & -18.83                        & -62.26                        & -41.70                        & -5.23                        & \bf -26.45                        \\
DCVC-FM~\shortcite{li2024neural}  & -19.27 & -16.72                        & -25.55                        & -42.46                        & -28.81                        & -24.26                        & \bf -26.19                        \\
VTM-23.0~\shortcite{bross2021overview}  & -40.21 & -35.59                        & -32.15                        & -38.05                        & -39.12                        & -41.19                        & \bf -37.72                        \\
%EEV-0.4  & 57.76 & 51.10                        & 23.29                        & 38.34                        & 43.84                        & 112.45                        & 68.05                        \\
EEV-0.5~\shortcite{deepye2024}  & -32.10                       & -13.55 &  -29.22 &  -61.76 &  -30.21 &  -34.92                        & \bf -33.63                        
\\
\bottomrule
\end{tabular}
\end{table*}

\begin{table*}[tb]
%\captionsetup{font=small}
\caption[]{BD-Rate(\%) performance comparison using MS-SSIM (RGB colorspace). The intra-period is set to be 32. Baseline is HM-16.20.}
\centering
%\footnotesize
\label{tab:1}
\begin{tabular}{c cccccc cc}
\toprule
  & \footnotesize{JCT-VC Class B} & \footnotesize{JCT-VC Class C} & \footnotesize{JCT-VC Class D} & \footnotesize{JCT-VC Class E} & \footnotesize{UVG} & \footnotesize{MCL-JCV} & \footnotesize{\bf Average} \\
\midrule
DCVC-DC~\shortcite{li2023neural}  & -39.71 & -43.94                        & -52.76                        & -10.68                        & -30.05                        & -33.73                        & \bf -35.15                        \\
DCVC-FM~\shortcite{li2024neural}  & -5.35 & -22.75                        & -22.99                        & -19.20                        & -3.42                        & -17.86                        & \bf -15.26                        \\
VTM-23.0~\shortcite{bross2021overview}  & -38.02 & -34.72                        & -31.00                        & -37.42                        & -38.77                        & -42.07                        & \bf -37.00                        \\
%EEV-0.4  & 57.76 & 51.10                        & 23.29                        & 38.34                        & 43.84                        & 112.45                        & 68.05                        \\
EEV-0.5~\shortcite{deepye2024}  & -45.49                        &  -38.36 &  -49.67 &  -54.33 &  -33.20 &  -28.17                        & \bf -41.53                        
\\
\bottomrule
\end{tabular}
\end{table*}

\subsection{Edge-computing Devices Oriented LVC}
The integration of hardware acceleration into LVC has emerged as a promising direction for achieving real-time and energy-efficient video coding. A novel pipeline, termed FPX-NVC (FPGA Accelerated Neural Video Coding), was proposed in ~\cite{jia2022fpx} to leverage edge-computing FPGA devices for accelerating neural operations in both intra and inter coding. The framework introduces two key contributions: an innovative network structure and an energy-efficient network deployment method. The associated intelligent video coding system was also established, composed of visual capturing, neural encoding, decoding, and display, realizing 4K ultra-high-definition (UHD) video content. 

Furthermore, the deployment of learned image compression (LIC) on hardware platforms, particularly field-programmable gate arrays (FPGAs), has gained significant attention due to its potential to achieve real-time processing while maintaining computational efficiency. A fine-grained pipeline architecture was proposed in ~\cite{sun2022f} to realize high digital signal processing (DSP) efficiency, leveraging 8-bit fixed-point arithmetic for network quantization. In summary, the FPGA devices accelerated LVC schemes significantly reduce the computational complexity, maintaining the model upgradeability and flexibility. Video services such as cloud-based video production and transcoding might be critical utility scenarios.

However, the disadvantages of the edge-computing acceleration for neural codecs are also obvious. The widely used neural inference engine usually supports standard operations such as convolution, softmax, ReLU-based activation, etc. Regarding the latest advanced operations, such as transformer and multi-head attention, the embedded support within those devices is limited. Therefore it is necessary to develop a more flexible LVC deployment mechanism.

\subsection{Mobile Devices Oriented LVC}
In the rapidly evolving field of neural video decoding, significant strides have been made to enhance real-time performance and efficiency on mobile devices. ~\cite{le2022mobilecodec}, proposed by Qualcomm, was the first inter-frame neural video decoder with an efficient network architecture running on a commercial mobile phone, decompressing 720P high-definition videos in real-time while maintaining a low bitrate and high visual quality. The special design of MobileCodec was that the separate quantization parameters for each output channel were employed for the convolutional weights, resulting in channel-wise quantization aware tuning. As its successor, the MobileNVC model~\cite{van2024mobilenvc} realized 1080P YUV420 video real-time decoding on a mobile device. The concepts of overlapped block motion compensation were enjoyed to reduce coding artifacts using the mobile neural accelerator. By doing so, the run-time memory consumption could also be reduced compared to frame-level motion compensation.

The neural inference ability on mobile phones has been exponentially increasing along with the advancement of smart phone hardware development. General-purpose neural inference engines are the building blocks and essential hardware support for LVC deployment on smartphones.

\subsection{Semantic Communications Systems}
Recent advancements in semantic communications have introduced innovative deep learning-based frameworks to enhance data transmission efficiency. ~\cite{9791398} proposes Nonlinear Transform Source-Channel Coding (NTSCC), which integrates nonlinear transform coding with deep joint source-channel coding to learn latent representations and entropy models, particularly excelling in high-resolution image transmission. ~\cite{zhang2024advance} critically reviews the advancements in semantic information and semantic communications, including theory, architecture, and potential applications. Moreover, the work deeply explores the key challenges in developing semantic communications and present the development prospects, aiming to prompt further scientific and industrial advances in semantic communications. %Meanwhile, \cite{10817394} introduces D²-JSCC, a digital deep joint source-channel coding framework that employs adaptive prior models and digital channel coding to achieve robust image transmission. These frameworks highlight the potential of deep learning in advancing semantic communication systems, with future research focusing on E2E metrics, multi-user extensions, and advanced channel techniques.

\section{Experiments of LVC Models}
\begin{figure*}[tb]
  \centering
  \footnotesize
  \subfloat[JCT-VC Class B]{
    \includegraphics[height=3.9cm]{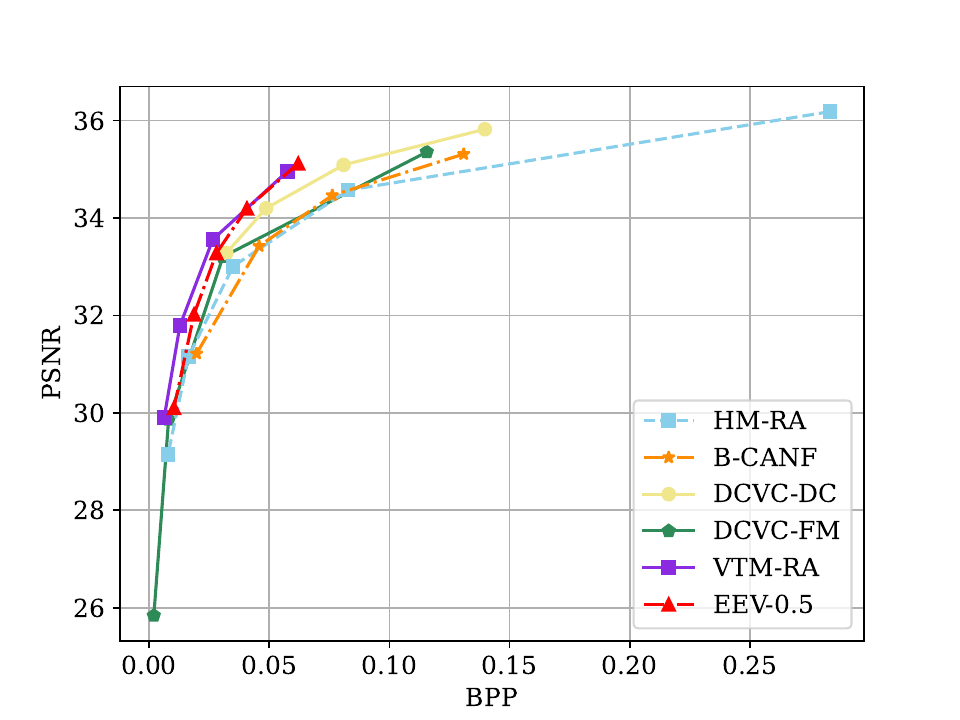}}\hfil
  \subfloat[JCT-VC Class D]{
    \includegraphics[height=3.9cm]{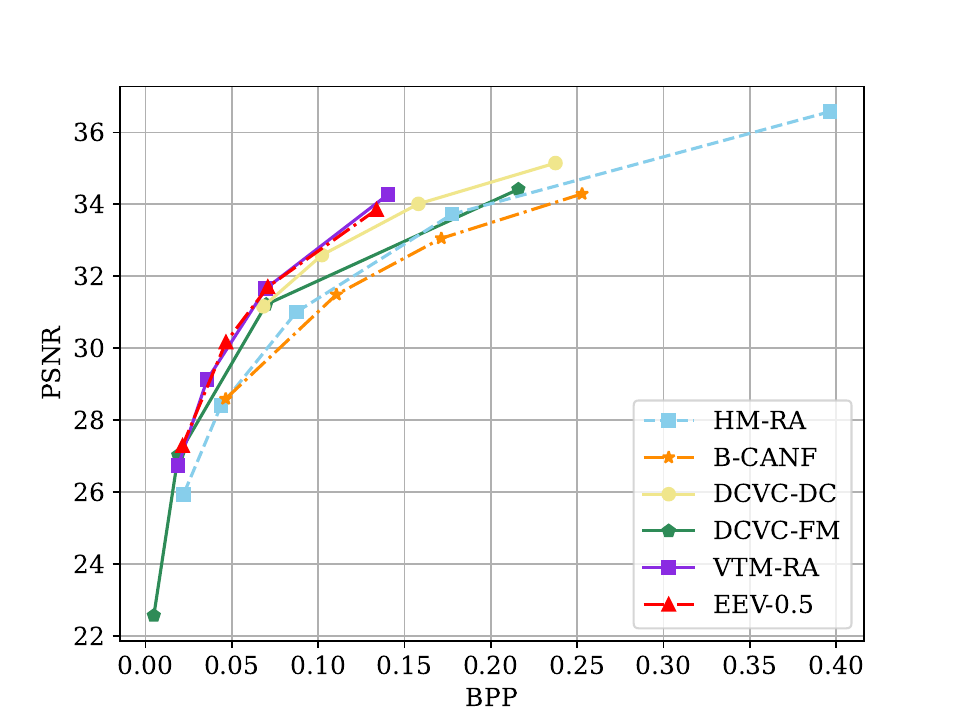}}\hfil
  \subfloat[JCT-VC Class E]{
    \includegraphics[height=3.9cm]{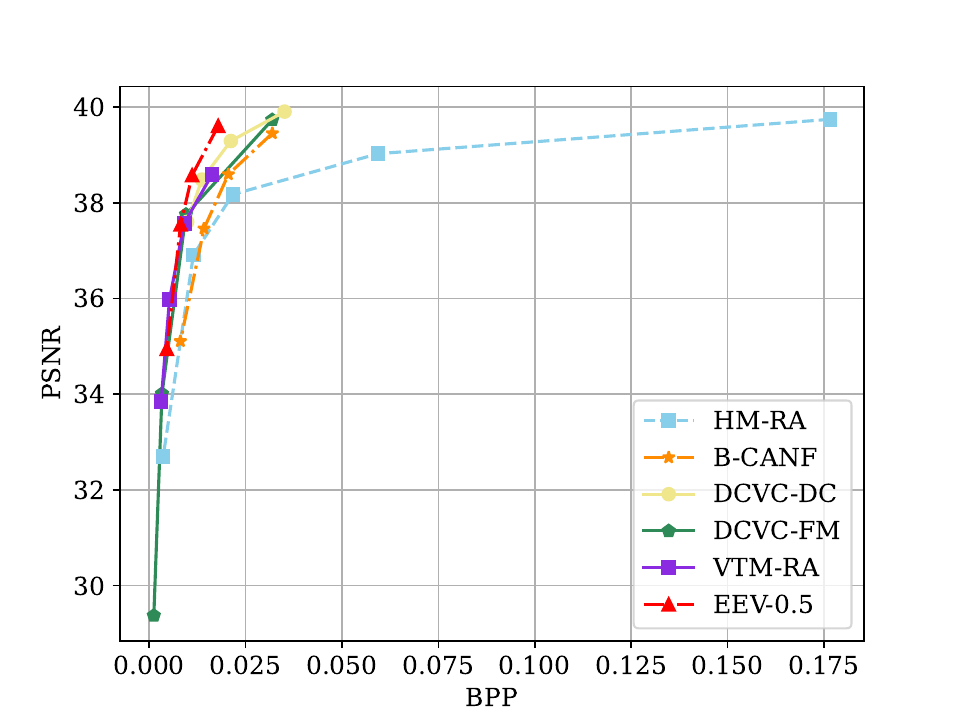}}
%\captionsetup{font=small}
  \caption{Rate-distortion curves for PSNR metric (RGB colorspace). HM/VTM indicates HEVC/VVC standards respectively. Four advanced LVC models are compared. (a) Class B ($1920\times1080$). (b) Class D ($416\times240$). (c) Class E ($1280\times720$). Note that LVCs perform extremely better than conventional codecs in Class E while competitive performances in other classes.}
  \label{fig:rd_curve}
\end{figure*}

{\bf Test Datasets. }
Performance evaluation datasets contain JCT-VC test sequences\cite{hevcdataset}, UVG dataset\cite{uvgdatasets} and MCL-JCV dataset\cite{mcljcv}, which are widely adopted in video compression research to measure the performance of coding models and algorithms. These test sequences include a diverse set of video clips with varying resolutions, frame rates, and content types, ensuring comprehensive testing under different scenarios. 

%{\bf Test configurations.}

{\bf Performance Evaluation Metrics. }
PSNR and MS-SSIM are two widely used metrics for evaluating video quality. PSNR measures the logarithmic ratio of the maximum pixel intensity to the Mean Squared Error (MSE) between original and compressed frames. MS-SSIM is a perceptually motivated metric that aligns more closely with the human subjective perception.

BD-rate is a widely used metric in video compression to evaluate the efficiency of coding algorithms. It quantifies the average bitrate savings or overhead of one method compared to another while maintaining the same visual quality. A negative BD-rate indicates bitrate savings, meaning the evaluated method achieves the same quality at a lower bitrate, while a positive BD-rate suggests higher bitrate requirements. 

\textbf{Test Conditions.} We evaluate our coding performances using the first 96 frames of each sequence. 
Our benchmarks include traditional video coding standards H.265/HEVC and H.266/VVC with random access configuration. Their reference model HM-16.20 and VTM-23.0 are selected. Regarding HM, we employ the default \textit{encoder\_randomaccess\_main\_rext} configuration with intra-period=32 and GOP-size=16. As for VTM, the default \textit{encoder\_randomaccess\_vtm} configuration is chosen with intra-period=32 and GOP-size=32. To achieve a better compression ratio for the traditional codecs, we follow the test conditions in \cite{li2023neural} to convert RGB videos to YUV444 to employ coding. The reconstructed YUV444 videos will be converted back to RGB for distortion calculation.

For LVC models, the intra-period and GOP sizes are set to the same value of 32. 
For uni-directional prediction NVC benchmark, DCVC-DC\cite{li2023neural} and DCVC-FM~\cite{li2024neural} are selected. For bi-directional NVC, EEV-0.5~\cite{deepye2024} is evaluated. MPAI-EEV is a standard project focused on developing an AI-based End-to-End (E2E) video coding standard, aiming to enhance video compression by utilizing AI technologies, moving beyond traditional data coding methods. Since 2022, MPAI-EEV has released 5 verification models and a public benchmark for unmanned-aerial-vehicles (UAV) video coding~\cite{jia2023mpai}.

\textbf{Experimental Results.} 
For uni-directional predictive NVC models, including DCVC-DC and DCVC-FM, it can be observed that there is a noticeable performance gap compared to the traditional coding method VTM in terms of PSNR metric. In contrast, bi-directional predictive NVC models exhibit slightly lower PSNR performance than VTM but significantly outperform VTM in MS-SSIM, highlighting their superior perceptual quality. However, the performance of uni-directional predictive NVC models may be improved using larger GOP sizes. In low-latency coding scenario, uni-directional predictive NVC models demonstrate their strengths and achieve competitive performance.

\section{Conclusion}
This paper presents a comprehensive overview of recent progress in end-to-end optimized intelligent video coding models and their associated system-level developments. We systematically review both uni-directional and bi-directional prediction based LVC methods, detailing their architectural designs and optimization strategies. Additionally, we explore the utilization of LVC in practical systems using resource-constrained and energy-efficient platforms, further enhancing the versatility of LVC. We also present experimental results comparing its performance with traditional codecs, demonstrating significant improvements in compression efficiency. Through this analysis, we aim to illustrate the potential of LVC in the future of video coding and its applicability across diverse hardware environments.

% \appendix

%\section*{Ethical Statement}

%There are no ethical issues.

% \section*{Acknowledgments}

% The preparation of these instructions and the \LaTeX{} and Bib\TeX{}
% files that implement them was supported by Schlumberger Palo Alto
% Research, AT\&T Bell Laboratories, and Morgan Kaufmann Publishers.
% Preparation of the Microsoft Word file was supported by IJCAI.  An
% early version of this document was created by Shirley Jowell and Peter
% F. Patel-Schneider.  It was subsequently modified by Jennifer
% Ballentine, Thomas Dean, Bernhard Nebel, Daniel Pagenstecher,
% Kurt Steinkraus, Toby Walsh, Carles Sierra, Marc Pujol-Gonzalez,
% Francisco Cruz-Mencia and Edith Elkind.

%% The file named.bst is a bibliography style file for BibTeX 0.99c
\bibliographystyle{named}
\bibliography{ijcai25}

\begin{thebibliography}{}

\bibitem[\protect\citeauthoryear{Alexandre \bgroup \em et al.\egroup }{2023}]{Alexandre2023HierarchicalBV}
David Alexandre, Hsueh-Ming Hang, and Wen-Hsiao Peng.
\newblock Hierarchical b-frame video coding using two-layer canf without motion coding.
\newblock In {\em CVPR}, 2023.

\bibitem[\protect\citeauthoryear{Ball{\'e} \bgroup \em et al.\egroup }{2018}]{balle2018integer}
Johannes Ball{\'e}, Nick Johnston, and David Minnen.
\newblock Integer networks for data compression with latent-variable models.
\newblock In {\em ICLR}, 2018.

\bibitem[\protect\citeauthoryear{Boyce \bgroup \em et al.\egroup }{2018}]{hevcdataset}
Jill Boyce, Karsten Suehring, Xiang Li, and Vadim Seregin.
\newblock Jvet-j1010: Jvet common test conditions and software reference configurations, 2018.

\bibitem[\protect\citeauthoryear{Bross \bgroup \em et al.\egroup }{2021}]{bross2021overview}
Benjamin Bross, Ye-Kui Wang, Yan Ye, Shan Liu, Jianle Chen, Gary~J Sullivan, and Jens-Rainer Ohm.
\newblock Overview of the versatile video coding (vvc) standard and its applications.
\newblock {\em IEEE TCSVT}, 2021.

\bibitem[\protect\citeauthoryear{Chen \bgroup \em et al.\egroup }{2017}]{chen2017deepcoder}
Tong Chen, Haojie Liu, Qiu Shen, Tao Yue, Xun Cao, and Zhan Ma.
\newblock Deepcoder: A deep neural network based video compression.
\newblock In {\em VCIP}, 2017.

\bibitem[\protect\citeauthoryear{Chen \bgroup \em et al.\egroup }{2022}]{chen2022learning}
Meixu Chen, Todd Goodall, Anjul Patney, and Alan~C Bovik.
\newblock Learning to compress videos without computing motion.
\newblock {\em SPIC}, 2022.

\bibitem[\protect\citeauthoryear{Chen \bgroup \em et al.\egroup }{2023a}]{chen2023sparse}
Jiancong Chen, Meng Wang, Pingping Zhang, Shurun Wang, and Shiqi Wang.
\newblock Sparse-to-dense: High efficiency rate control for end-to-end scale-adaptive video coding.
\newblock {\em IEEE TCSVT}, 2023.

\bibitem[\protect\citeauthoryear{Chen \bgroup \em et al.\egroup }{2023b}]{Chen2022BCANFAB}
Mu-Jung Chen, Yi-Hsin Chen, and Wen-Hsiao Peng.
\newblock B-canf: Adaptive b-frame coding with conditional augmented normalizing flows.
\newblock {\em IEEE TCSVT}, 2023.

\bibitem[\protect\citeauthoryear{Cui \bgroup \em et al.\egroup }{2021}]{cui2021asymmetric}
Ze~Cui, Jing Wang, Shangyin Gao, Tiansheng Guo, Yihui Feng, and Bo~Bai.
\newblock Asymmetric gained deep image compression with continuous rate adaptation.
\newblock In {\em CVPR}, 2021.

\bibitem[\protect\citeauthoryear{Dai \bgroup \em et al.\egroup }{2022}]{9791398}
Jincheng Dai, Sixian Wang, Kailin Tan, Zhongwei Si, Xiaoqi Qin, Kai Niu, and Ping Zhang.
\newblock Nonlinear transform source-channel coding for semantic communications.
\newblock {\em IEEE JSAC}, 2022.

\bibitem[\protect\citeauthoryear{Djelouah \bgroup \em et al.\egroup }{2019}]{djelouah2019neural}
Abdelaziz Djelouah, Joaquim Campos, Simone Schaub-Meyer, and Christopher Schroers.
\newblock Neural inter-frame compression for video coding.
\newblock In {\em ICCV}, 2019.

\bibitem[\protect\citeauthoryear{Fathima \bgroup \em et al.\egroup }{2023}]{fathima2023neural}
Noor Fathima, Jens Petersen, Guillaume Sauti{\`e}re, Auke Wiggers, and Reza Pourreza.
\newblock A neural video codec with spatial rate-distortion control.
\newblock In {\em WACV}, 2023.

\bibitem[\protect\citeauthoryear{Feng \bgroup \em et al.\egroup }{2020}]{Feng2020LearnedVC}
Runsen Feng, Yaojun Wu, Zongyu Guo, Zhizheng Zhang, and Zhibo Chen.
\newblock Learned video compression with feature-level residuals.
\newblock In {\em CVPRW}, 2020.

\bibitem[\protect\citeauthoryear{Feng \bgroup \em et al.\egroup }{2021}]{Feng2021VersatileLV}
Runsen Feng, Zongyu Guo, Zhizheng Zhang, and Zhibo Chen.
\newblock Versatile learned video compression.
\newblock {\em arXiv preprint: 2111.03386}, 2021.

\bibitem[\protect\citeauthoryear{Ho \bgroup \em et al.\egroup }{2022}]{Ho2022CANFVCCA}
Yung-Han Ho, Chih-Peng Chang, Peng-Yu Chen, Alessandro Gnutti, and Wen-Hsiao Peng.
\newblock Canf-vc: Conditional augmented normalizing flows for video compression.
\newblock In {\em ECCV}, 2022.

\bibitem[\protect\citeauthoryear{Hong \bgroup \em et al.\egroup }{2020}]{hong2020efficient}
Weixin Hong, Tong Chen, Ming Lu, Shiliang Pu, and Zhan Ma.
\newblock Efficient neural image decoding via fixed-point inference.
\newblock {\em IEEE TCSVT}, 2020.

\bibitem[\protect\citeauthoryear{Hu \bgroup \em et al.\egroup }{2020}]{hu2020improving}
Zhihao Hu, Zhenghao Chen, Dong Xu, Guo Lu, Wanli Ouyang, and Shuhang Gu.
\newblock Improving deep video compression by resolution-adaptive flow coding.
\newblock In {\em ECCV}, 2020.

\bibitem[\protect\citeauthoryear{Hu \bgroup \em et al.\egroup }{2021}]{hu2021fvc}
Zhihao Hu, Guo Lu, and Dong Xu.
\newblock Fvc: A new framework towards deep video compression in feature space.
\newblock In {\em CVPR}, 2021.

\bibitem[\protect\citeauthoryear{Hu \bgroup \em et al.\egroup }{2022}]{9880063}
Zhihao Hu, Guo Lu, Jinyang Guo, Shan Liu, Wei Jiang, and Dong Xu.
\newblock Coarse-to-fine deep video coding with hyperprior-guided mode prediction.
\newblock In {\em CVPR}, 2022.

\bibitem[\protect\citeauthoryear{Jia \bgroup \em et al.\egroup }{2022}]{jia2022fpx}
Chuanmin Jia, Xinyu Hang, et~al.
\newblock Fpx-nic: An fpga-accelerated 4k ultra-high-definition neural video coding system.
\newblock {\em IEEE TCSVT}, 2022.

\bibitem[\protect\citeauthoryear{Jia \bgroup \em et al.\egroup }{2023}]{jia2023mpai}
Chuanmin Jia, Feng Ye, Fanke Dong, Kai Lin, Leonardo Chiariglione, et~al.
\newblock Mpai-eev: Standardization efforts of artificial intelligence based end-to-end video coding.
\newblock {\em IEEE TCSVT}, 2023.

\bibitem[\protect\citeauthoryear{Kim \bgroup \em et al.\egroup }{2023}]{Kim2023NeuralVC}
Yeong-Chun Kim, Suyong Bahk, Seung~Hwan Kim, Won~Hee Lee, Dokwan Oh, and Hui-Yong Kim.
\newblock Neural video compression with temporal layer-adaptive hierarchical b-frame coding.
\newblock {\em arXiv preprint: 2308.15791}, 2023.

\bibitem[\protect\citeauthoryear{Ladune \bgroup \em et al.\egroup }{2020}]{ladune2020optical}
Th{\'e}o Ladune, Pierrick Philippe, Wassim Hamidouche, Lu~Zhang, and Olivier D{\'e}forges.
\newblock Optical flow and mode selection for learning-based video coding.
\newblock In {\em MMSP}, 2020.

\bibitem[\protect\citeauthoryear{Ladune \bgroup \em et al.\egroup }{2021}]{Ladune2021ConditionalCF}
Th{\'e}o Ladune, Pierrick Philippe, Wassim Hamidouche, Lu~Zhang, and Olivier D{\'e}forges.
\newblock Conditional coding for flexible learned video compression.
\newblock In {\em ICLRW}, 2021.

\bibitem[\protect\citeauthoryear{Le \bgroup \em et al.\egroup }{2022}]{le2022mobilecodec}
Hoang Le, Liang Zhang, Amir Said, Guillaume Sautiere, Yang Yang, Pranav Shrestha, Fei Yin, Reza Pourreza, and Auke Wiggers.
\newblock Mobilecodec: neural inter-frame video compression on mobile devices.
\newblock In {\em ACM MMSys}, 2022.

\bibitem[\protect\citeauthoryear{Li \bgroup \em et al.\egroup }{2021}]{li2021deep}
Jiahao Li, Bin Li, and Yan Lu.
\newblock Deep contextual video compression.
\newblock In {\em NeurIPS}, 2021.

\bibitem[\protect\citeauthoryear{Li \bgroup \em et al.\egroup }{2022}]{Li_2022}
Jiahao Li, Bin Li, and Yan Lu.
\newblock Hybrid spatial-temporal entropy modelling for neural video compression.
\newblock ACM MM, 2022.

\bibitem[\protect\citeauthoryear{Li \bgroup \em et al.\egroup }{2023}]{li2023neural}
Jiahao Li, Bin Li, and Yan Lu.
\newblock Neural video compression with diverse contexts.
\newblock In {\em CVPR}, 2023.

\bibitem[\protect\citeauthoryear{Li \bgroup \em et al.\egroup }{2024}]{li2024neural}
Jiahao Li, Bin Li, and Yan Lu.
\newblock Neural video compression with feature modulation.
\newblock In {\em CVPR}, 2024.

\bibitem[\protect\citeauthoryear{Lin \bgroup \em et al.\egroup }{2020}]{lin2020m}
Jianping Lin, Dong Liu, Houqiang Li, and Feng Wu.
\newblock M-lvc: Multiple frames prediction for learned video compression.
\newblock In {\em CVPR}, 2020.

\bibitem[\protect\citeauthoryear{Lin \bgroup \em et al.\egroup }{2021}]{lin2021modulated}
Jianping Lin, Dong Liu, Jie Liang, Houqiang Li, and Feng Wu.
\newblock Modulated variable-rate deep video compression.
\newblock In {\em DCC}, 2021.

\bibitem[\protect\citeauthoryear{Lin \bgroup \em et al.\egroup }{2022}]{lin2022content}
Chih-Hsuan Lin, Yi-Hsin Chen, and Wen-Hsiao Peng.
\newblock Content-adaptive motion rate adaption for learned video compression.
\newblock In {\em IEEE PCS}, 2022.

\bibitem[\protect\citeauthoryear{Liu \bgroup \em et al.\egroup }{2020}]{liu2020deep}
Dong Liu, Yue Li, Jianping Lin, Houqiang Li, and Feng Wu.
\newblock Deep learning-based video coding: A review and a case study.
\newblock {\em ACM CSUR}, 2020.

\bibitem[\protect\citeauthoryear{Liu \bgroup \em et al.\egroup }{2021}]{liu2020neural}
Haojie Liu, Ming Lu, Zhan Ma, Fan Wang, Zhihuang Xie, Xun Cao, and Yao Wang.
\newblock Neural video coding using multiscale motion compensation and spatiotemporal context model.
\newblock {\em IEEE TCSVT}, 2021.

\bibitem[\protect\citeauthoryear{Lu \bgroup \em et al.\egroup }{2019}]{lu2019dvc}
Guo Lu, Wanli Ouyang, Dong Xu, Xiaoyun Zhang, Chunlei Cai, and Zhiyong Gao.
\newblock Dvc: An end-to-end deep video compression framework.
\newblock In {\em CVPR}, 2019.

\bibitem[\protect\citeauthoryear{Lu \bgroup \em et al.\egroup }{2020}]{lu2020content}
Guo Lu, Chunlei Cai, Xiaoyun Zhang, et~al.
\newblock Content adaptive and error propagation aware deep video compression.
\newblock In {\em ECCV}, 2020.

\bibitem[\protect\citeauthoryear{Lu \bgroup \em et al.\egroup }{2021}]{9072487}
Guo Lu, Xiaoyun Zhang, Wanli Ouyang, Li~Chen, Zhiyong Gao, and Dong Xu.
\newblock An end-to-end learning framework for video compression.
\newblock {\em IEEE TPAMI}, 2021.

\bibitem[\protect\citeauthoryear{Ma \bgroup \em et al.\egroup }{2019}]{ma2019image}
Siwei Ma, Xinfeng Zhang, Chuanmin Jia, Zhenghui Zhao, Shiqi Wang, and Shanshe Wang.
\newblock Image and video compression with neural networks: A review.
\newblock {\em IEEE TCSVT}, 2019.

\bibitem[\protect\citeauthoryear{Ma \bgroup \em et al.\egroup }{2022}]{ma2022evolution}
Siwei Ma, Li~Zhang, Shiqi Wang, Chuanmin Jia, Shanshe Wang, Tiejun Huang, Feng Wu, and Wen Gao.
\newblock Evolution of avs video coding standards: twenty years of innovation and development.
\newblock {\em Science China Information Sciences}, 2022.

\bibitem[\protect\citeauthoryear{Mao \bgroup \em et al.\egroup }{2021}]{mao2020neural}
Hongzi Mao, Chenjie Gu, Miaosen Wang, Angie Chen, Nevena Lazic, Nir Levine, et~al.
\newblock Neural rate control for video encoding using imitation learning.
\newblock In {\em ICMLW}, 2021.

\bibitem[\protect\citeauthoryear{Mercat \bgroup \em et al.\egroup }{2020}]{uvgdatasets}
Alexandre Mercat, Marko Viitanen, and Jarno Vanne.
\newblock Uvg dataset: 50/120fps 4k sequences for video codec analysis and development.
\newblock In {\em ACM MMSys}, 2020.

\bibitem[\protect\citeauthoryear{Pergament \bgroup \em et al.\egroup }{2022}]{pergament2022pim}
Evgenya Pergament, Pulkit Tandon, Oren Rippel, et~al.
\newblock Pim: Video coding using perceptual importance maps.
\newblock {\em arXiv preprint: 2212.10674}, 2022.

\bibitem[\protect\citeauthoryear{Pourreza and Cohen}{2021}]{pourreza2021extending}
Reza Pourreza and Taco Cohen.
\newblock Extending neural p-frame codecs for b-frame coding.
\newblock In {\em ICCV}, 2021.

\bibitem[\protect\citeauthoryear{Rippel \bgroup \em et al.\egroup }{2021}]{rippel2021elf}
Oren Rippel, Alexander~G Anderson, Kedar Tatwawadi, Sanjay Nair, Craig Lytle, and Lubomir Bourdev.
\newblock Elf-vc: Efficient learned flexible-rate video coding.
\newblock In {\em ICCV}, 2021.

\bibitem[\protect\citeauthoryear{Sheng \bgroup \em et al.\egroup }{2022}]{sheng2021temporal}
Xihua Sheng, Jiahao Li, Bin Li, Li~Li, Dong Liu, and Yan Lu.
\newblock Temporal context mining for learned video compression.
\newblock {\em IEEE TMM}, 2022.

\bibitem[\protect\citeauthoryear{Sheng \bgroup \em et al.\egroup }{2024a}]{sheng2024spatial}
Xihua Sheng, Li~Li, Dong Liu, and Houqiang Li.
\newblock Spatial decomposition and temporal fusion based inter prediction for learned video compression.
\newblock {\em IEEE TCSVT}, 2024.

\bibitem[\protect\citeauthoryear{Sheng \bgroup \em et al.\egroup }{2024b}]{sheng2024bi}
Xihua Sheng, Li~Li, Dong Liu, and Shiqi Wang.
\newblock Bi-directional deep contextual video compression.
\newblock {\em arXiv preprint arXiv:2408.08604}, 2024.

\bibitem[\protect\citeauthoryear{Shi \bgroup \em et al.\egroup }{2022}]{Shi2022AlphaVCHA}
Yibo Shi, Yunying Ge, Jing Wang, and Jue Mao.
\newblock Alphavc: High-performance and efficient learned video compression.
\newblock In {\em ECCV}, 2022.

\bibitem[\protect\citeauthoryear{Sun \bgroup \em et al.\egroup }{2022}]{sun2022f}
Heming Sun, Qingyang Yi, Fangzheng Lin, Lu~Yu, Jiro Katto, and Masahiro Fujita.
\newblock F-lic: Fpga-based learned image compression with a fine-grained pipeline.
\newblock In {\em IEEE A-SSCC}, 2022.

\bibitem[\protect\citeauthoryear{Van~Rozendaal \bgroup \em et al.\egroup }{2024}]{van2024mobilenvc}
Ties Van~Rozendaal, Tushar Singhal, Hoang Le, Guillaume Sautiere, Amir Said, Krishna Buska, et~al.
\newblock Mobilenvc: Real-time 1080p neural video compression on a mobile device.
\newblock In {\em WACV}, 2024.

\bibitem[\protect\citeauthoryear{Veerabadran \bgroup \em et al.\egroup }{2020}]{veerabadran2020adversarial}
Vijay Veerabadran, Reza Pourreza, et~al.
\newblock Adversarial distortion for learned video compression.
\newblock In {\em CVPRW}, 2020.

\bibitem[\protect\citeauthoryear{Wang \bgroup \em et al.\egroup }{2016}]{mcljcv}
Haiqiang Wang, Weihao Gan, Sudeng Hu, Joe~Yuchieh Lin, Lina Jin, Longguang Song, Ping Wang, Ioannis Katsavounidis, Anne Aaron, and C-C~Jay Kuo.
\newblock Mcl-jcv: a jnd-based h. 264/avc video quality assessment dataset.
\newblock In {\em IEEE ICIP}, 2016.

\bibitem[\protect\citeauthoryear{Wu \bgroup \em et al.\egroup }{2018}]{wu2018video}
Chao-Yuan Wu, Nayan Singhal, and Philipp Krahenbuhl.
\newblock Video compression through image interpolation.
\newblock In {\em ECCV}, 2018.

\bibitem[\protect\citeauthoryear{Xu \bgroup \em et al.\egroup }{2023}]{xu2023bit}
Tongda Xu, Han Gao, Chenjian Gao, Yuanyuan Wang, Dailan He, Jinyong Pi, Jixiang Luo, Ziyu Zhu, Mao Ye, Hongwei Qin, et~al.
\newblock Bit allocation using optimization.
\newblock In {\em ICML}, 2023.

\bibitem[\protect\citeauthoryear{Yang \bgroup \em et al.\egroup }{2020}]{yang2020learning}
Ren Yang, Fabian Mentzer, Luc~Van Gool, and Radu Timofte.
\newblock Learning for video compression with hierarchical quality and recurrent enhancement.
\newblock In {\em CVPR}, 2020.

\bibitem[\protect\citeauthoryear{Yang \bgroup \em et al.\egroup }{2022}]{yang2021perceptual}
Ren Yang, Radu Timofte, and Luc Van~Gool.
\newblock Perceptual learned video compression with recurrent conditional gan.
\newblock In {\em IJCAI}, 2022.

\bibitem[\protect\citeauthoryear{Ye \bgroup \em et al.\egroup }{2024}]{deepye2024}
Feng Ye, Li~Zhang, and Chuanmin Jia.
\newblock Deep video compression with scaled hierarchical bi-directional motion model.
\newblock In {\em ACM MM}, 2024.

\bibitem[\protect\citeauthoryear{Y{\i}lmaz and Tekalp}{2021}]{yilmaz2021end}
M~Ak{\i}n Y{\i}lmaz and A~Murat Tekalp.
\newblock End-to-end rate-distortion optimized learned hierarchical bi-directional video compression.
\newblock {\em IEEE TIP}, 2021.

\bibitem[\protect\citeauthoryear{Zhang \bgroup \em et al.\egroup }{2024a}]{zhang2024advance}
Ping Zhang, Yiming Liu, Yile Song, and Jiaxiang Zhang.
\newblock Advances and challenges in semantic communications: A systematic review.
\newblock {\em National Science Open}, 2024.

\bibitem[\protect\citeauthoryear{Zhang \bgroup \em et al.\egroup }{2024b}]{zhang2024neural}
Yiwei Zhang, Guo Lu, Yunuo Chen, Shen Wang, Yibo Shi, Jing Wang, and Li~Song.
\newblock Neural rate control for learned video compression.
\newblock In {\em ICLR}, 2024.

\end{thebibliography}

\end{document}